\documentclass[proceedings]{JHEP3}

\newcommand{\cQ}{{\cal Q}}
\newcommand{\Bx}{x_{\rm B}}
\newcommand{\GeV}{\mbox{GeV}}

\PrHEP{PrHEP hep2001}                 
\conference{International Europhysics Conference on HEP}   

\usepackage{epsfig}                   

\title{Overview of deeply virtual Compton scattering}

\author{A.V.~Belitsky\thanks{This work, UMD-PP\#02-018, was supported by the US DOE
                             under contract DE-FG02-93ER40762.} \\  
        Department of Physics, University of Maryland,
        MD 20742-4111, College Park, USA \\                         
        E-mail: \email{belitsky@physics.umd.edu}}

\author{\speaker{D.~M\"uller} \\                                    
        Fachbereich Physik, Universit\"at  Wuppertal,
        D-42097 Wuppertal, Germany \\                               
        E-mail: \email{dmueller@theorie.physik.uni-wuppertal.de}}   

\abstract{
We give a short overview of recent developments in understanding of the
deeply virtual Compton scattering on the proton target.
}

\begin{document}

\section{Introduction}

Deeply virtual Compton scattering (DVCS), recently measured by HERMES
\cite{Aip01}, H1 \cite{Adl01} and CLAS \cite{Ste01} collaborations at DESY
and Jefferson Lab, is the cleanest hadronic reaction that gives access to
the generalized parton distributions (GPDs) \cite{GPDs}. GPDs are the
probability amplitudes to knock out a parton from, say, the nucleon and to
put it back with a different longitudinal momentum fraction. The probed
partonic state is characterized, of course, by quantum numbers, like flavor,
spin, etc. GPDs unify a number of known concepts of hadronic physics: they
are related to parton densities, wave functions, and parton form factors,
coupled also to higher spin probes. Recall that conventional form factors
are encoded in matrix elements of local currents without derivatives, while
for the case at hand, operators contain any number $n$ of derivatives. For
$n = 1$ the form factors carry information on the orbital angular momentum
carried by constituents in the proton, see the second paper of Ref.\
\cite{GPDs}. Apart from DVCS and other related two-photon processes, GPDs also
contribute to the hard leptoproduction of mesons, however, their theoretical
description involves a new unknown nonperturbative input, a meson
distributions amplitude, which complicates the disentanglement of GPDs from
measurements.

\section{Differential cross section of $e N \to e' N' \gamma$}
\label{SecAziAngDep}

Presently, we concentrate on the structure of the cross section for
electroproduction of the real photon off the proton $e (k) N (P_1) \to e
(k^\prime) N (P_2) \gamma (q_2)$, whose amplitude is the sum of Bethe-Heitler
(BH) and the wanted DVCS amplitude. The four-fold cross section
\begin{eqnarray}
\label{WQ}
\frac{d\sigma}{d\Bx dy d|\Delta^2| d\phi}
=
\frac{\alpha^3  \Bx y } { 8 \, \pi \,  {\cal Q}^2}
\left( 1 + \frac{4 M^2 \Bx^2}{{\cal Q}^2} \right)^{-1/2}
\left|
\frac{
{\cal T}_{\rm BH} + {\cal T}_{\rm DVCS}
}{
e^3
}
\right|^2 \, ,
\end{eqnarray}
depends on the Bjorken variable $\Bx = {\cal Q}^2/(2 P_1 \cdot q_1)$, the squared
$t$-channel momentum transfer $\Delta^2 = (P_2 - P_1)^2$, the lepton energy loss
$y = P_1 \cdot q_1/ P_1 \cdot k$, and the azimuthal angle $\phi$, see Fig.\
\ref{kinematics}. The resolution scale $\cQ^2 = - q_1^2$ is given by the virtuality
of the incoming photon. In our frame, the angular dependence of separate components
in the cross section (\ref{WQ}) is given by a finite sum of Fourier harmonics:
\begin{eqnarray}
\label{Par-BH}
&&|{\cal T}_{\rm BH}|^2
= \frac{e^6}
{\Bx^2 y^2  \Delta^2\, {\cal P}_1 (\phi) {\cal P}_2 (\phi)}
\left\{
c^{\rm BH}_0
+  \sum_{n = 1}^2
\left[ c^{\rm BH}_n \,\cos{(n\phi)} + s^{\rm BH}_n \,\sin{(n\phi)}
\right]
\right\}
\, , \\
\label{AmplitudesSquared}
&& |{\cal T}_{\rm DVCS}|^2
=
\frac{e^6}{y^2 {\cal Q}^2}\left\{
c^{\rm DVCS}_0
+ \sum_{n=1}^2
\left[
c^{\rm DVCS}_n \cos (n\phi) + s^{\rm DVCS}_n \sin (n \phi)
\right]
\right\}
\, , \nonumber\\
\label{InterferenceTerm}
&&
{\cal T}_{\rm DVCS} {\cal T}_{\rm BH}^\ast
+ {\cal T}_{\rm DVCS}^\ast {\cal T}_{\rm BH}
= \frac{\pm e^6}{\Bx y^3 {\cal P}_1 (\phi) {\cal P}_2 (\phi) \Delta^2}
\left\{
c_0^{\cal I}
+ \sum_{n = 1}^3
\left[
c_n^{\cal I} \cos(n \phi) +  s_n^{\cal I} \sin(n \phi)
\right]
\right\}
\, . \nonumber
\end{eqnarray}

\FIGURE[h]{\mbox{\epsfxsize=7cm \epsfbox{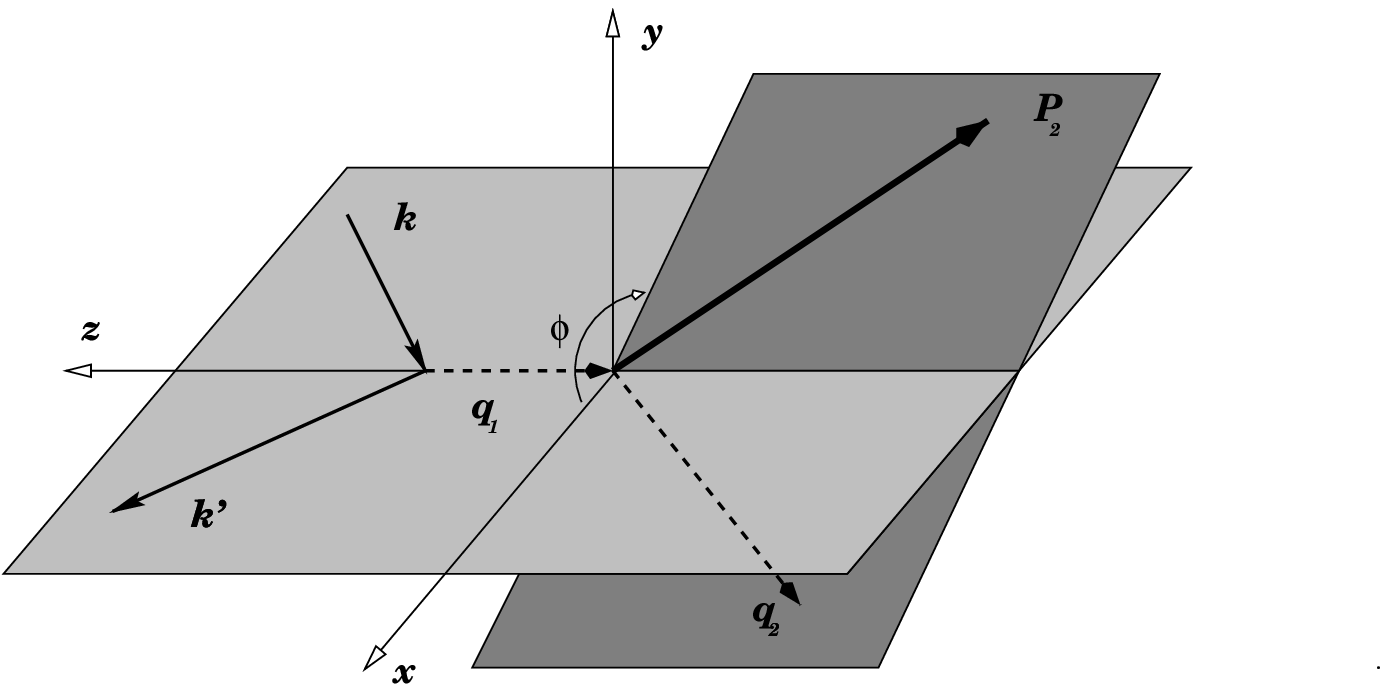}}
\caption{
\label{kinematics}
The  kinematics for leptoproduction in the target rest frame, where the
azimuthal of the proton momentum with respect to the lepton scattering
plane is $\phi$.
}}

\noindent
Here the $+$ ($-$) sign stands for electron (po\-si\-tron) beam. In the squared
BH and the interference term an additional $\phi$ dependence arises from the
scaled lepton BH propagators ${\cal P}_i (\phi) = A_i + B_i \cos(\phi) $, which
can be rather strong for large $y$. The coefficients $c_n$ and $s_n$ depend on
the kinematical variables $y$, $\Bx$, $\Delta^2$, ${\cal Q}^2$, and lepton and
hadron polarizations.

The Fourier coefficients (FCs) $c^{\cal I}_1$ and $s^{\cal I}_1$ [$c^{\cal I}_0$,
$c^{\cal I}_2$, and $s^{\cal I}_2$] as well as $c^{\rm DVCS}_0$ [$c^{\rm DVCS}_1$,
and $s^{\rm DVCS}_1$] appear at the twist-two [-three] level, while even higher
harmonics are suppressed by powers of $\alpha_s$ or $1/\cQ^2$. For an unpolarized
target we have in the twist-two sector
\begin{eqnarray}
c^{\rm DVCS}_{0,{\rm unp}}
&=&
2 ( 2 - 2 y + y^2 )
{\cal C}^{\rm DVCS}_{\rm unp}
\left(
{\cal F},{\cal F}^\ast\right), \\
\label{IntIm}
\left\{{c^{\cal I}_{1, \rm unp} \atop s^{\cal I}_{1, \rm unp}}\right\}
&=&
8 K
\left\{  {-(2 - 2y + y^2) \atop \lambda y (2-y)} \right\}
\left\{{{\rm Re} \atop {\rm Im} } \right\}
{\cal C}^{\cal I}_{\rm unp}\left({\cal F} \right) \, ,
\end{eqnarray}
where $K \approx \sqrt{(1 - \Bx)(1 - y)} \Delta_\perp/\cQ$ and $\lambda$ is the
lepton helicity. The ${\cal C}$s are functions of the so-called Compton form
factors (CFFs) ${\cal F} = \{{\cal H}, {\cal E}, \widetilde {\cal H}, \widetilde
{\cal E}\}$ which parametrize the DVCS tensor and they have been worked out in
Refs.\ \cite{DVCStw2} at twist-two level, e.g.,
\begin{eqnarray*}
&&{\cal C}^{\cal I}_{\rm unp}
=
F_1 {\cal H}
+ \frac{\Bx}{2 - \Bx} (F_1 + F_2) \widetilde {\cal H}
- \frac{\Delta^2}{4M^2} F_2 {\cal E}
\, , \qquad
{\cal C}^{\rm DVCS}_{\rm unp}
=
\frac{1}{(2 - \Bx)^2}
\bigg\{
4 (1 - \Bx) | {\cal H} |^2
\nonumber\\
&&
+
4 (1 - \Bx)
| \widetilde {\cal H} |^2
-
2 \Bx^2 \Re{\rm e}
\left(
{\cal H} {\cal E}^\ast + \widetilde {\cal H} \widetilde {\cal E}^\ast
\right)
-
\left(
\Bx^2 + (2 - \Bx)^2 \frac{\Delta^2}{4 M^2}
\right)
| {\cal E} |^2
- \Bx^2 \frac{\Delta^2}{4 M^2}
| \widetilde {\cal E} |^2
\bigg\} \, ,
\end{eqnarray*}
where $F_1$ and $F_2$ are the nucleon Dirac and Pauli form factors, respectively.
The twist-three analysis \cite{DVCStw3} (see also \cite{GoePolVan01}) has just been
completed and will be presented in detail elsewhere. The CFFs are expressed in terms
of GPDs $F = \{ H, E, \widetilde H, \widetilde E \}$. At leading order (LO) of
perturbation theory the relations read
\begin{equation}
\label{GPDs2CFFs}
\left\{
{
{\cal H} \atop \widetilde {\cal H}
}
\right\} (\xi, \Delta^2, \cQ^2)
=
\sum_{i = u, d, s}
\int_{-1}^1 \frac{d x}{\xi}
\left(
\frac{Q^2_i}{1 - x/\xi - i 0}
\mp \{ \xi \to - \xi \}
\right)
\left\{
{
H_i \atop \widetilde H_i
}
\right\} (x, \xi, \Delta^2, \cQ^2) \, ,
\end{equation}
where $Q_i$ is the fractional quark charge and $\xi = \Bx/(2 - \Bx)$. Analogous
formulae are valid for the spin-flip CFFs ${\cal E}$ and $\widetilde {\cal E}$.
Unfortunately, from Eq.\ (\ref{GPDs2CFFs}) one can not practically deconvolute
GPDs. However, as Eq.\ (\ref{IntIm}) demonstrates, those observables which are
sensitive to $s^{\cal I}_1$ allow to measure directly GPDs (at LO) for $x = \pm
\xi$. Generally, one has to rely on models with a set of free parameters, which
have to be adjusted to experimental data.

\section{Observables and extraction of model parameters}
\label{SecEstvsDat}

For the low-${\cal Q}^2$ kinematical settings of the present experiments
\cite{Aip01,Ste01} it is desirable to separate the twist-two and -tree sector,
since a priori the latter can contaminate the former. This
procedure is practically possible only with facilities where positive and negative
lepton beams are available by means of charge asymmetries and Fourier analysis.

\FIGURE[h]{\epsfxsize=6.5cm \epsfbox{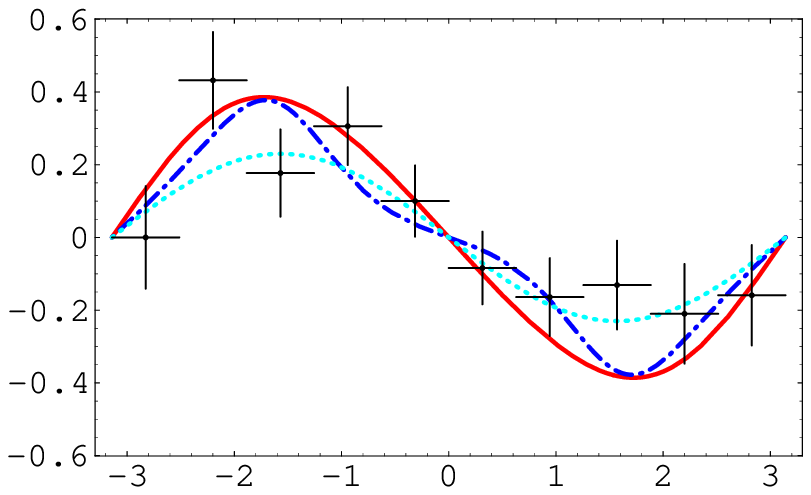}
\caption{
\label{HERMES}
$A_{\rm LU}$ as a function of $\phi^\prime_\gamma = \pi - \phi$ at HERMES
versus model predictions: A$_{WW}$ (solid) and B$_{WW+qGq}$ (dash-dotted).
}}

The charge-odd combinations of cross section (\ref{WQ}) gives the interference
term whose coefficients can be extracted by the weighting procedure,
\begin{equation}
\left\{
{ c_n^{\cal I} \atop s_n^{\cal I} }
\right\}
\propto
\int_0^{2\pi} d w
\left\{
{\cos \atop \sin}
\right\}
(n\phi)
\left(
\frac{d^+ \sigma }{d\phi} - \frac{d^- \sigma }{d\phi}
\right),
\end{equation}
with the measure $d w \propto {\cal P}_1 (\phi) {\cal P}_2 (\phi) d \phi$.
The FCs of the squared DVCS term can be obtained from the charge-even part
after subtraction of the squared BH term, i.e.\ $d^{\rm DVCS}\sigma = ( d \sigma^+
+ d^-\sigma)/2 - d^{\rm BH} \sigma$, and weighting with respect to the
measure $d \phi$. Asymmetries are normalized with respect to $\int_0^{2\pi}
d \phi \frac{d^+ \sigma + d^-\sigma}{d\phi}$, which does not contain
twist-three corrections. Unfortunately, for single-charge lepton beam
machines this clear separation of twist-two and -three effects cannot be
achieved -- also not with spin or azimuthal asymmetries.
For instance, for the beam-spin asymmetry measured with HERMES \cite{Aip01} and
CLAS \cite{Ste01},
\begin{equation}
\label{LUasym}
A_{\rm LU} (\phi) =
\left( d \sigma^\uparrow - d \sigma^\downarrow \right)
/
\left( d \sigma^\uparrow + d \sigma^\downarrow \right)
\, ,
\end{equation}
the leading twist term $s_1^{\cal I}$ in the numerator will be affected by
$s_2^{\cal I}$ and $s_1^{\rm DVCS}$ while the denominator will be the sum of
all three contributions in Eq.\ (\ref{Par-BH}). Fortunately, if the
condition $(1-y)\Delta^2/ y^2 \cQ^2 \ll 1$ is fulfilled, $s_1^{\rm DVCS}$ is
kinematically suppressed and, moreover, the squared BH term $c_0^{\rm BH}$
dominates in the denominator. Thus, the lepton BH propagators approximately
cancel in Eq.\ (\ref{LUasym}) and $A_{\rm LU} (\phi) \approx \pm (\Bx
s_1^{\cal I} /y c_0^{\rm BH}) \sin(\phi)$.

Since the number of experimental observables is very limited we are confined to
use GPD models in order to confront theoretical expectations with data. Let us
briefly outline, however, a way how model parameters can be fit to experiments in
particular kinematical situations. An ansatz is based on a simplistic
factorization of $\Delta^2$ and $(x, \xi)$ dependence. Sum rules and reduction
formulae and ad hoc assumptions result into a model for, e.g., $H_i$
\begin{eqnarray}
\label{Hansatz}
H_i = F_1^i (\Delta^2)
\int_{0}^1 d y \int_{- 1 + y}^{1 - y} d z \,
\delta ( y + \xi z - x ) \, q_i ( y) \, N (b_i)
\frac{\left[ (1 - y)^2 - z^2 \right]^{b_i}}{(1 - y)^{2 b_i + 1}}
+ D_i
\, ,
\end{eqnarray}
where $q_i$ are the parton densities given at the input scale $\cQ_0$ and $N
(b_i) = \frac{{\mit\Gamma} \left( b_i + 3/2 \right)} {\sqrt{\pi}
{\mit\Gamma} (b_i + 1)}$. For valence quarks the form factors are fixed by
the sum rules, while for sea quarks we parametrize them as $F_1^{\rm sea
}(\Delta^2) = \left( 1 - B_{\rm sea} \Delta^2 / 3 \right)^{- 3}$ with a free
slope parameter $B_{\rm sea}$. The free parameters $b_i$ in Eq.\
(\ref{Hansatz}) control the skewedness effects. The last term in
(\ref{Hansatz}) stands for the mesonic-like contributions, see
\cite{GoePolVan01}. Parametrizations for the spin-flip GPDs are obtained by
the replacements: $F_1^i \to F_2^i$ and $D_i \to -D_i$. $\widetilde H_i$ is
given in terms of polarized parton densities, while $\widetilde E_i$ is
dominated by the pion pole, see \cite{GoePolVan01}.

\FIGURE[h]{\epsfxsize=6.5cm \epsfbox{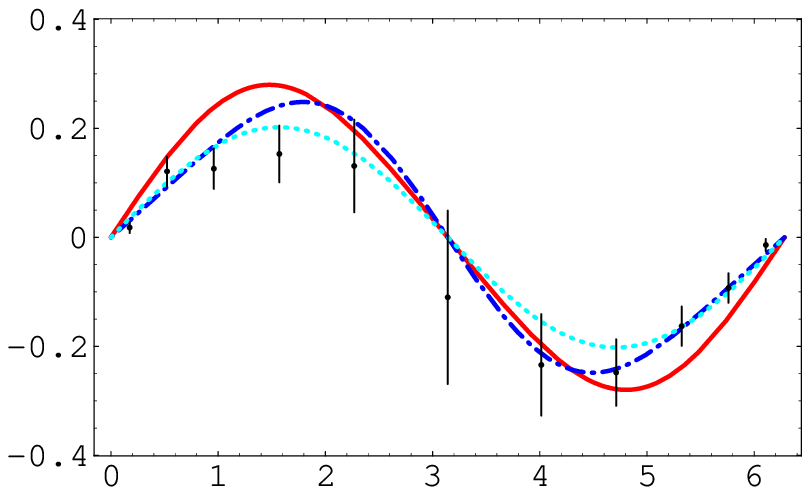}
\caption{
\label{CLAS}
$A_{\rm LU}$ as a function of $\phi^\prime_\gamma = \pi - \phi$ at CLAS
versus model predictions. Same assignments as in Fig.\ \ref{HERMES}.
}}

Assuming the MRS A$'$ and GS A densities at the input scale $\cQ_0^2=4\
\mbox{GeV}^2$, we analyzed the existing DVCS data with these models at LO in
the twist-three approximation with neglected ${\cal Q}^2$-evolution. The
unpolarized cross section at small $\Bx$ as measured by H1 \cite{Adl01} is
dominated by ${\cal H}_{\rm sea}$, i.e., ${\cal C}^{\rm DVCS}_{\rm unp}
\approx |{\cal H}|^2$ for $-\Delta^2 \ll 4 M^2$. We found that the skewedness
effect is small and the fall-off with $\Delta^2$ is strong. The choice
$b_{\rm sea} \to \infty$ and $B_{\rm sea} = 9 \ \GeV^2$ gives a good agreement
with the data. Using $b_{\rm val}$ and sea-quark magnetic moment $\kappa_{\rm sea}$
as free parameters we distinguish two models: A with $b_{\rm val} = 1$ and
$\kappa_{\rm sea} = 0$, and B with $b_{\rm val} \to \infty$ and $\kappa_{\rm sea}
= - 3$.

In Fig.\ \ref{HERMES} and \ref{CLAS} we give our predictions for the model
A with twist-three GPDs taken in the Wandzura-Wilczek approximation (solid
line) \cite{DVCStw3} and the model B with quark-gluon-quark correlations
accounted for in the twist-three GPDs (dash-dotted line) versus experimental
data. For HERMES and CLAS plots, the dotted line denotes the $\sin \phi$ with
the amplitudes $0.23$ and $0.202$, respectively, fit to the data by the
collaborations. At HERMES the integrated beam-spin asymmetry $A_{\rm LU}  = - 0.23
 \pm 0.04 ({\rm stat}) \pm 0.03 ({\rm syst})$ has
been measured at the average values $\langle \cQ^2 \rangle = 2.6\ \GeV^2$,
$\langle \Bx \rangle = 0.11$ and $\langle - \Delta^2 \rangle = 0.27\ \GeV^2$
\cite{Aip01}. Integrating our predictions A$_{WW}$ (solid) and B$_{WW+qGq}$
(dash-dotted) from Fig.\ \ref{HERMES} we have $A_{\rm LU} = - 0.27$ and
$A_{\rm LU} = - 0.16$, respectively. Both of them are compatible with the
data. The beam-spin asymmetry as measured by CLAS collaboration was integrated
over the region: $1\ \GeV^2 < \cQ^2 < 1.75\ \GeV^2 $, $0.13 < \Bx < 0.35$,
and $0.1\ \GeV^2 < - \Delta^2 < 0.3\ \GeV^2$ with $W > 2\ \GeV$. Their
integrated result is fit by $A_{\rm LU}(\phi) = \alpha \sin(\phi) + \beta
\sin(2\phi)$ with $\alpha = 0.202 \pm 0.028 ({\rm stat}) \pm 0.013 ({\rm syst})$
and $\beta = -0.024 \pm 0.021 ({\rm stat}) \pm 0.009 ({\rm syst})$ \cite{Ste01}.
The model A$_{WW}$ fails to describe the data resulting into $\alpha = 0.28$
and $\beta = 0.014$. However for the model B$_{WW+qGq}$ we find $\alpha = 0.24$
and $\beta = - 0.03$ in fairly good agreement with the measurement. Note that
contrary to $\alpha$, the second harmonic parametrized by $\beta$ is very
sensitive to quark-gluon-quark correlations.

\section{Beyond leading order and power}
\label{SecBeyLO}

So far we have shown that the DVCS results measured in three different
experiments are consistent for a set of GPD models with DVCS amplitudes
evaluated at LO within the twist-three approximation. Once we will have a
rough idea on the magnitude of model parameters a refined analysis has to
be conducted taking into account higher order effects in QCD coupling constant.
In the twist-two sector, they are completely worked out at NLO \cite{NLOcor}.
According to our recent analysis, their size is moderate for each quark species,
however, large radiative corrections can be generated by certain models of
gluon GPDs. The latter are induced by strong evolution effects and, thus, these
models can be experimentally tested in the small-$\Bx$ region. The gluonic
contributions can be set to zero by an appropriate choice of the factorization
scale. In this scheme we have moderate $20\%$ corrections to the CFFs, however,
the DVCS cross section then gets corrected by $40\%$.

As we have observed the CLAS result presumably indicates a non-negligible
contribution from multiparton correlations in the nucleon. Therefore, a
further study of higher twist effects is necessary. Since $\cQ^2$ is of the
order of one $\GeV^2$, one expects also a sizable contamination by target mass
$M^2/{\cal Q}^2$ effects. A first step in this direction has been made in
\cite{BelMue01a}, however, a complete treatment requires a consideration of
mass effects stemming form multiparticle operators.


\end{document}